\documentclass{webofc}
\usepackage[varg]{txfonts}   
\usepackage{slashed}

\begin{document}
\title{Scattering Interference effects on $H^+\to t\bar b$ Signals in MSSM Benchmark Scenarios}

\author{\firstname{Abdesslam} \lastname{Arhrib}\inst{1}\fnsep
    \and
        \firstname{Duarte} \lastname{Azevedo}\inst{2}\fnsep
             \and
        \firstname{Rachid} \lastname{Benbrik}\inst{3}\fnsep
             \and
        \firstname{Hicham} \lastname{Harouiz}\inst{3}\fnsep
             \and
        \firstname{Stefano} \lastname{Moretti}\inst{4}\fnsep
             \and
        \firstname{Riley} \lastname{Patrick}\inst{5}\fnsep
             \and
        \firstname{Rui} \lastname{Santos}\inst{2,6}\fnsep
}

\institute{Faculty of Sciences and Techniques, Abdelmalek Essaadi University, B.P. 416, Tanger, Morocco 
\and
           Centro de F\'{\i}sica Te\'{o}rica e Computacional,     Faculdade de Ci\^{e}ncias, Universidade de Lisboa, Campo Grande, Edif\'{\i}cio C8,
  1749-016 Lisboa, Portugal 
\and
           MSISM Team, Facult\'e Polydisciplinaire de Safi, Sidi Bouzid, B.P. 4162, Safi, Morocco
\and
		   School of Physics and Astronomy, University of Southampton, Southampton, SO17 1BJ, United Kingdom 
\and
		   ARC Center of Excellence for Particle Physics at the Terascale, Department of Physics, University of Adelaide, 5005 Adelaide, South
Australia
\and
		   ISEL - Instituto Superior de Engenharia de Lisboa, Instituto Polit\'ecnico de Lisboa,  1959-007 Lisboa, Portugal
          }

\abstract{%
  In this talk an investigation into the interference effects between the process $pp\to \bar{t}bH^+$ followed by the decay $H^+ \to t\bar{b}$ and the background process $pp \to t\bar{t}b\bar{b}$ is presented. The level of interference in parts of the parameter space is shown to be high and as such it may spoil the results of typical analyses which treat signal and background as independent. This is shown for two benchmarks of the MSSM. 
}
\maketitle
\section{Introduction}
\label{intro}

The selection of benchmark is important as the chosen parameter point must provide both a non-negligible cross section and high width-to-mass ratio of the signal particle - in this case the charged Higgs. Two areas of the MSSM parameter space of interest to experimentalists are the hMSSM and the $m_{h}^{\rm mod+}$ models, for reviews of these see Refs.~\cite{Djouadi:2013uqa} and \cite{Heinemeyer:1999zf} respectively. We utilize these models as a vehicle for study of interference effects and hone in further on specific choices of MSSM parameters in the next section.

\section{Benchmark Selection}

To generate our benchmark points a scan of the MSSM parameter space is undertaken using \texttt{FeynHiggs}~\cite{Heinemeyer:1998yj}\cite{Hahn:2009zz} interfaced with \texttt{HiggsBounds-5.2.0beta}~\cite{Bechtle:2008jh}\cite{Bechtle:2011sb}\cite{Bechtle:2013wla}\cite{Bechtle:2015pma} and \texttt{HiggsSignals-2.2.0beta}~\cite{Bechtle:2013xfa} to obtain constraints.

\begin{figure}
\centering
\includegraphics[scale=0.30]{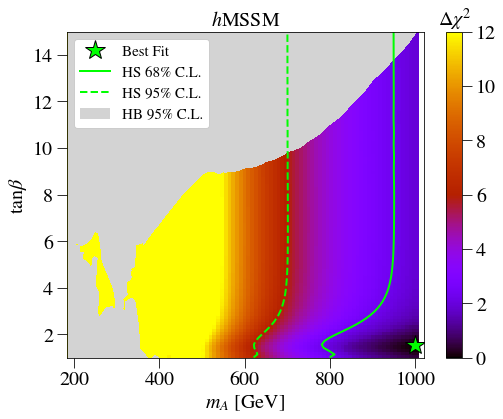}
\includegraphics[scale=0.30]{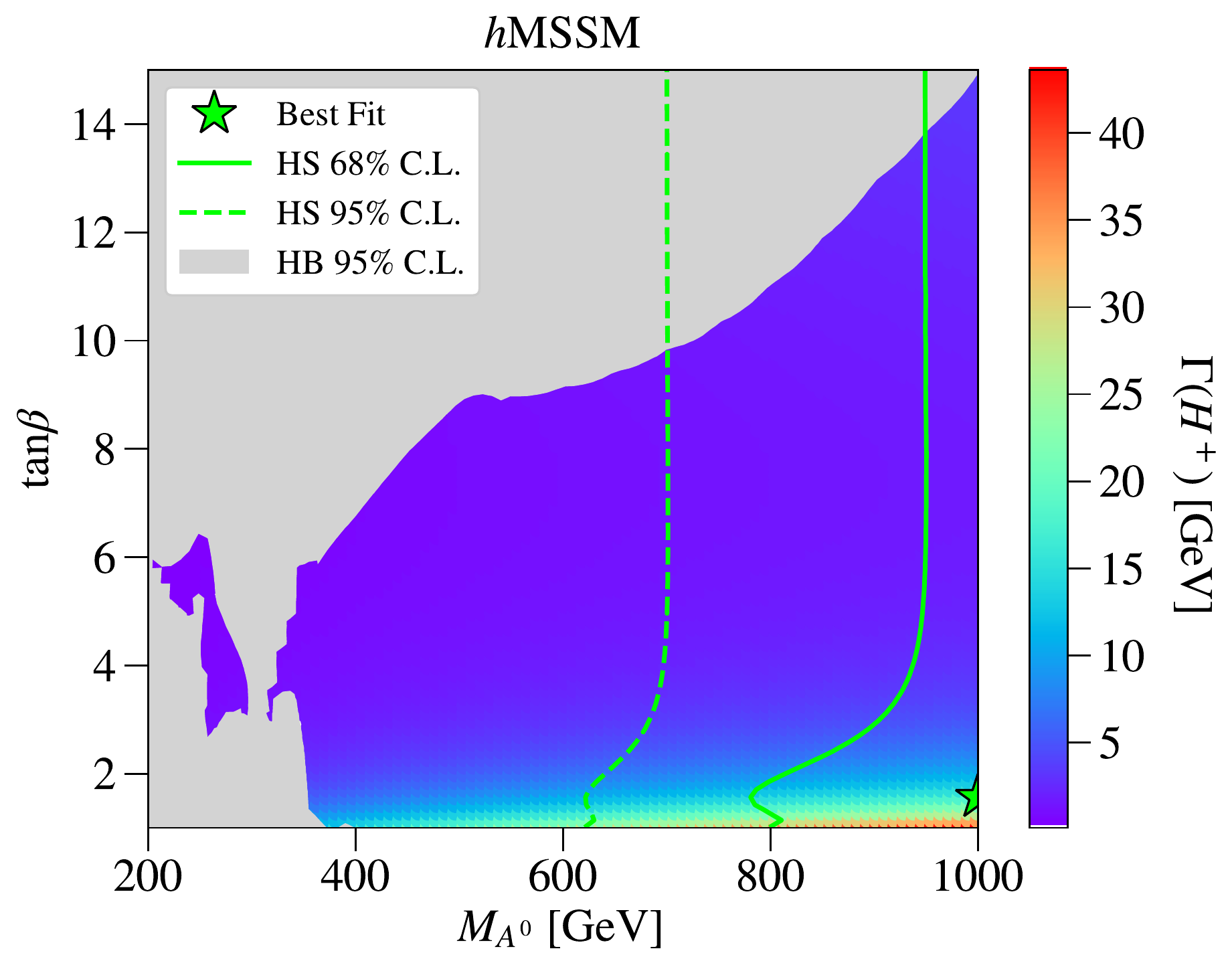}    \\
\includegraphics[scale=0.30]{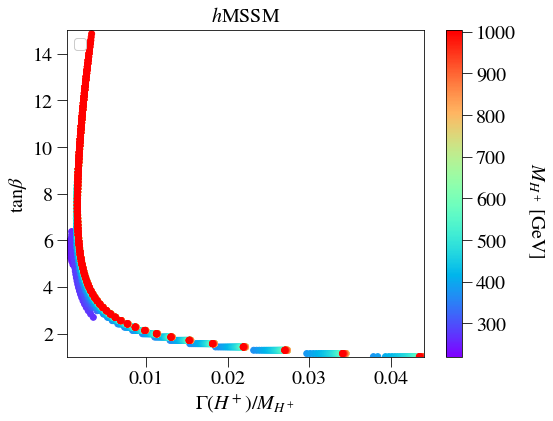}
\includegraphics[scale=0.30]{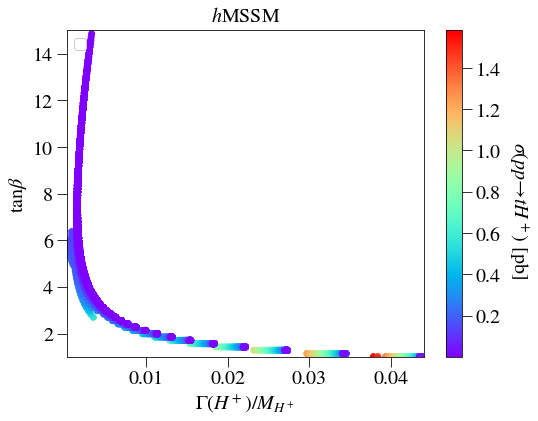}
\caption{
$\Delta\chi^2$ (top-left) and the charged Higgs total width (top-right) in the ($m_A\equiv M_{A^0}$, $\tan\beta$) plane. The best fit point is located at $M_{A^0}\approx 1$ TeV and $\tan\beta\approx 2$.  
The green lines show the exclusion limits from HiggSignals at $1\sigma$ (solid) and $2\sigma$ (dashed) while the gray 
area is ruled out by the various LHC searches implemented in HiggsBounds. The ratio $\Gamma_{H^\pm}/M_{H^\pm}$ as a function of the charged Higgs mass 
is shown in the bottom-left panel while in the bottom-right one it is presented as a function of the charged Higgs production cross section. }
\label{fig:hMSSM_HB5HS2_excl}
\end{figure}

Fig~\ref{fig:hMSSM_HB5HS2_excl} displays various three dimensional slices of the parameter space for the hMSSM model. We wish to choose a point which has high width-to-mass ratio, so as to generate high levels of interference, as well as a non-negligible cross section and high branching ratio of $H^+ \to t\bar{b}$ to preserve that cross section. Utilizing a scan of the parameter space with these goals in mind the best point found has $M_{H^{\pm}} = 633.91$ GeV and $\tan\beta = 1.01$.

\begin{figure}
\centering
\includegraphics[scale=0.30]{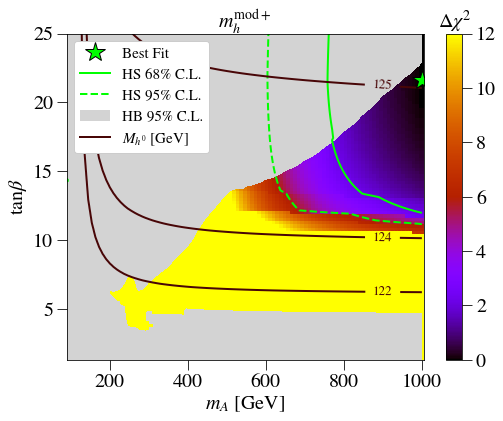}
\includegraphics[scale=0.30]{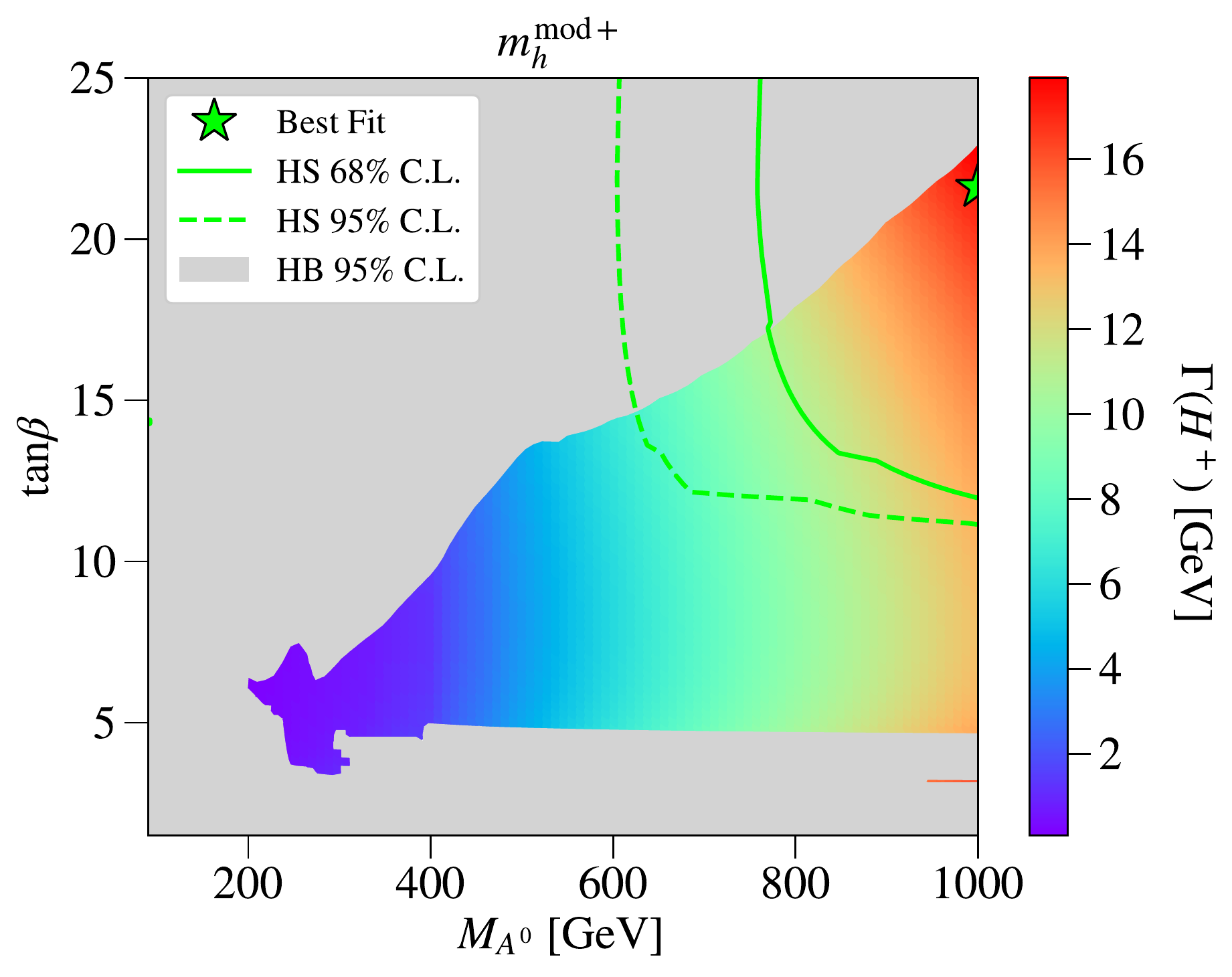}
\includegraphics[scale=0.30]{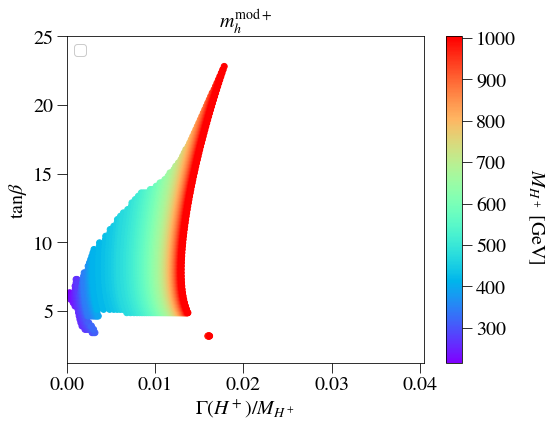}
\includegraphics[scale=0.30]{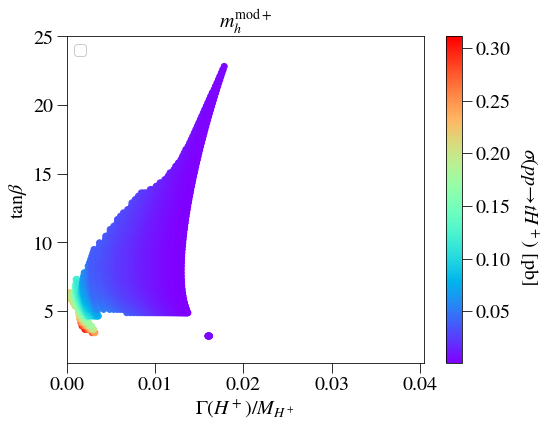}
\caption{
Allowed parameter region in the $m_{h}^{\rm mod+}$ scenario over the ($m_A\equiv M_{A^0}$, $\tan\beta$) plane with colour showing $\Delta\chi^2$ (top-left)
and the charged Higgs boson mass (top-right). The LHC Higgs searches constraints are included. The light green contours are HiggsSignals exclusion 
limits at $1\sigma$ (solid) and $2\sigma$ (dashed). The light gray area is excluded by HiggsBounds at $2\sigma$.
The solid brown lines are contours for the lighter CP-even scalar $h^0$ mass. The best fit point is located at $M_{H^\pm}\approx 1$ TeV and $\tan\beta =20$.
In the two bottom panels of Fig. \ref{fig:mhmodp_HB5HS2_excl} we present $\tan \beta$ as a function
of  $\Gamma_{H^\pm}/M_{H^\pm}$ with the colour code showing the charged Higgs mass (left) and
the charged Higgs production cross section (right).  }
\label{fig:mhmodp_HB5HS2_excl}
\end{figure}

Fig~\ref{fig:mhmodp_HB5HS2_excl} displays various three dimensional slices of the parameter space for the $m_{h}^{\rm mod+}$. In this case we choose a point which has a much lower width-to-mass ratio but a higher cross section and thus we may see lower levels of interference prior to cuts. It is possible however that interference may grow after cuts. Utilizing a scan of the parameter space with these goals in mind the best point found has $M_{H^{\pm}} = 303.08$ GeV and $\tan\beta = 3.42$.

\newpage

\section{Results}

For this analysis two independent samples were generated for both the hMSSM and $m_{h}^{\rm mod+}$ at leading order and at 13 TeV CoM energy at the LHC. The first sample contained 20,000,000 parton level events to provide a Monte Carlo sample with sufficiently low statistical error on the interference term. This sample was generated using \texttt{MadGraph}~\cite{Alwall:2014hca}. The second sample contained 500,000 events generated in \texttt{MadGraph}, then passed to \texttt{Pythia}~\cite{Sjostrand:2014zea} for hadronization/fragmentation and then to \texttt{Delphes}~\cite{deFavereau:2013fsa} for detector smearing.

\begin{table}[!ht]
		\centering
		\begin{tabular}{|lccccc|}
		\hline
	    {Model}  &  & {S} {(pb)} &  {B} {(pb)} &  {S+B} {(pb)} &  {I} {(pb)}  \\\hline
		{$h$MSSM}  & $\sigma$       & $0.032402$ & $13.078$ & $13.139$ & $0.028$ \\\hline
		 			    & $\Delta\sigma$ & $1.4\times 10^{-5}$ & $0.002$ & $0.001$ & $0.003$ \\\hline
		$m^{\rm{mod+}}_{{h}}$  & $\sigma$       & $0.088536$ & $13.095$ & $13.197$ & $0.014$ \\\hline
		                & $\Delta\sigma$ & $3.3\times 10^{-5}$ & $0.001$ & $0.001$ & $0.002$ \\\hline
		\end{tabular}
	\caption{\label{tab:partonresults} Parton level results for the $h$MSSM and $m^{\rm mod+}_{\rm h}$ benchmarks.}
\end{table}

This process was undertaken for the signal (``S''), defined as all processes in the MSSM that generate $t\bar{t}b\bar{b}$ mediated by $H^{\pm}$, the background (``B''), defined as all processes in the MSSM that generate $t\bar{t}b\bar{b}$ that are not mediated by $H^{\pm}$, and the signal plus background - from here out called total (``T'') - that contains all processes in the MSSM that can generate $t\bar{t}b\bar{b}$.

The three samples were then used to quantify the interference using the relationship $I = T - S - B$. As all three amplitudes have the same phase space then this equation isolates the interference term of the full scattering amplitude. The parton level results can be seen in Tab~\ref{tab:partonresults}.

\subsection{Event Reconstruction}

We choose the final state with exactly 1-lepton, thus allowing us to calculate the longitudinal momentum of missing energy. This is done via solving the following quadratic equation:
\begin{align}
	p_\nu^z = \frac{1}{2p_{\ell T}^2}\left(A_Wp^z_{\ell}\pm E_\ell\sqrt{A_W^2\pm 4p^2_{\ell T}E^2_{\nu T}}\right)
\end{align}
where, $A_W = M^2_{W^{\pm}} + 2p_{\ell T}\cdot E_{\nu T}$. If both of these solutions are non-real we veto the event.

We then perform a full event reconstruction by simultaneous minimisation of the following equations by permuting through all combinations of jets in the process and the two solutions for neutrino momentum,
				\begin{align}
				\chi^2_{\rm had} = \frac{\left(M_{\ell\nu} - M_W\right)^2}{\Gamma^2_W} 
									+\frac{\left(M_{jj} - M_W\right)^2}{\Gamma^2_W}		
									+\frac{\left(M_{\ell\nu j} - M_T\right)^2}{\Gamma^2_T}
								    +\frac{\left(M_{jjj} - M_T\right)^2}{\Gamma^2_T} 
									+\frac{\left(M_{jjjj} - M_{H^{\pm}}\right)^2}{\Gamma^2_{H^{\pm}}}
				\end{align} 
and
				\begin{align}
				\chi^2_{\rm lep} = \frac{\left(M_{\ell\nu} - M_W\right)^2}{\Gamma^2_W} 
									+\frac{\left(M_{jj} - M_W\right)^2}{\Gamma^2_W}		
									+\frac{\left(M_{\ell\nu j} - M_T\right)^2}{\Gamma^2_T}
								    +\frac{\left(M_{jjj} - M_T\right)^2}{\Gamma^2_T} 
									+\frac{\left(M_{\ell\nu jj} - M_{H^{\pm}}\right)^2}{\Gamma^2_{H^{\pm}}}
				\end{align}
  
\begin{figure}[!ht]
\centering
\includegraphics[scale=0.3]{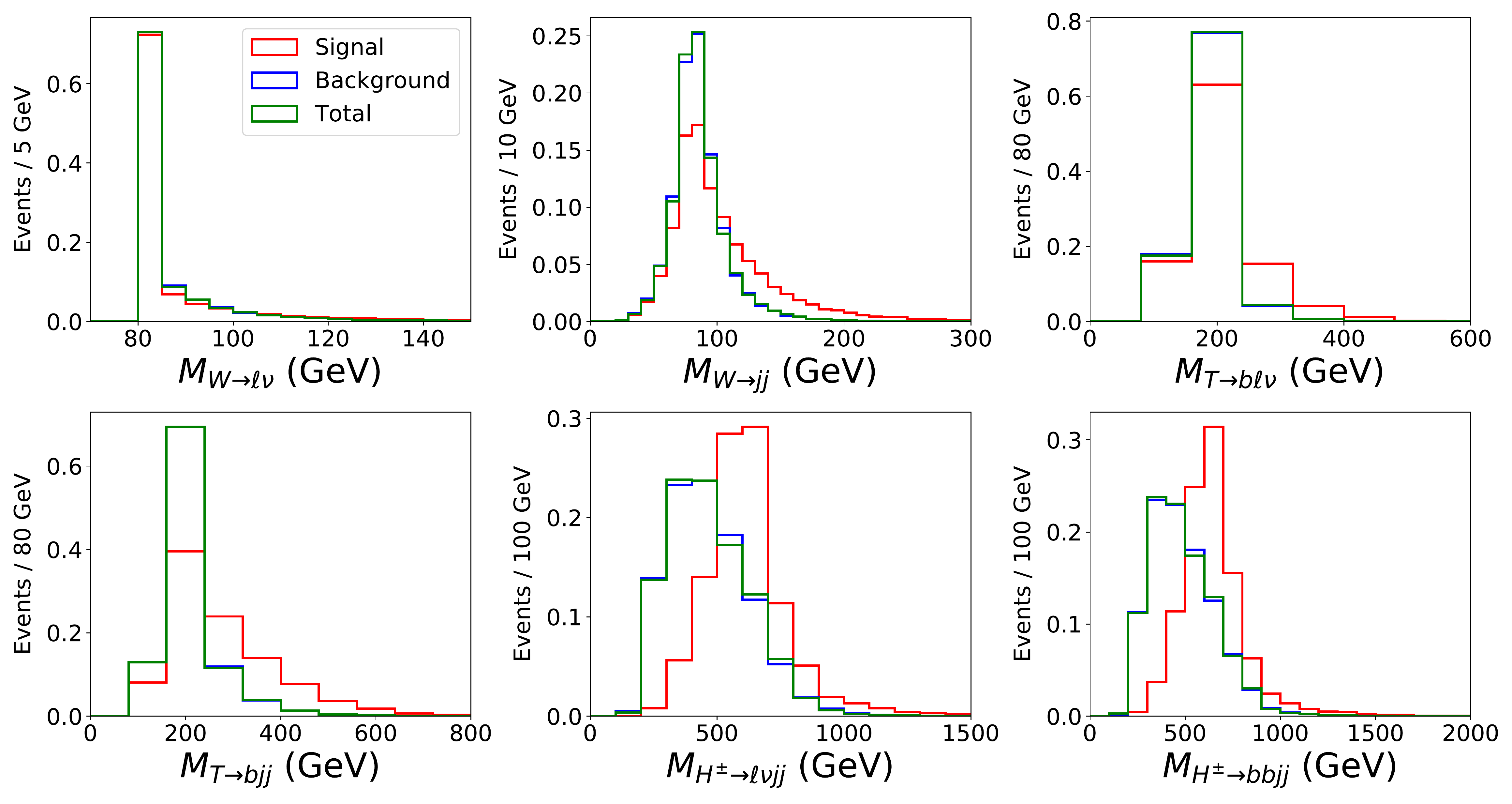}
\caption{\label{fig:hmssm-recon-variables} Invariant mass distributions for reconstructed particles in the $h$MSSM benchmark.}
\end{figure}
\begin{figure}[!ht]
	\centering
	\includegraphics[scale=0.3]{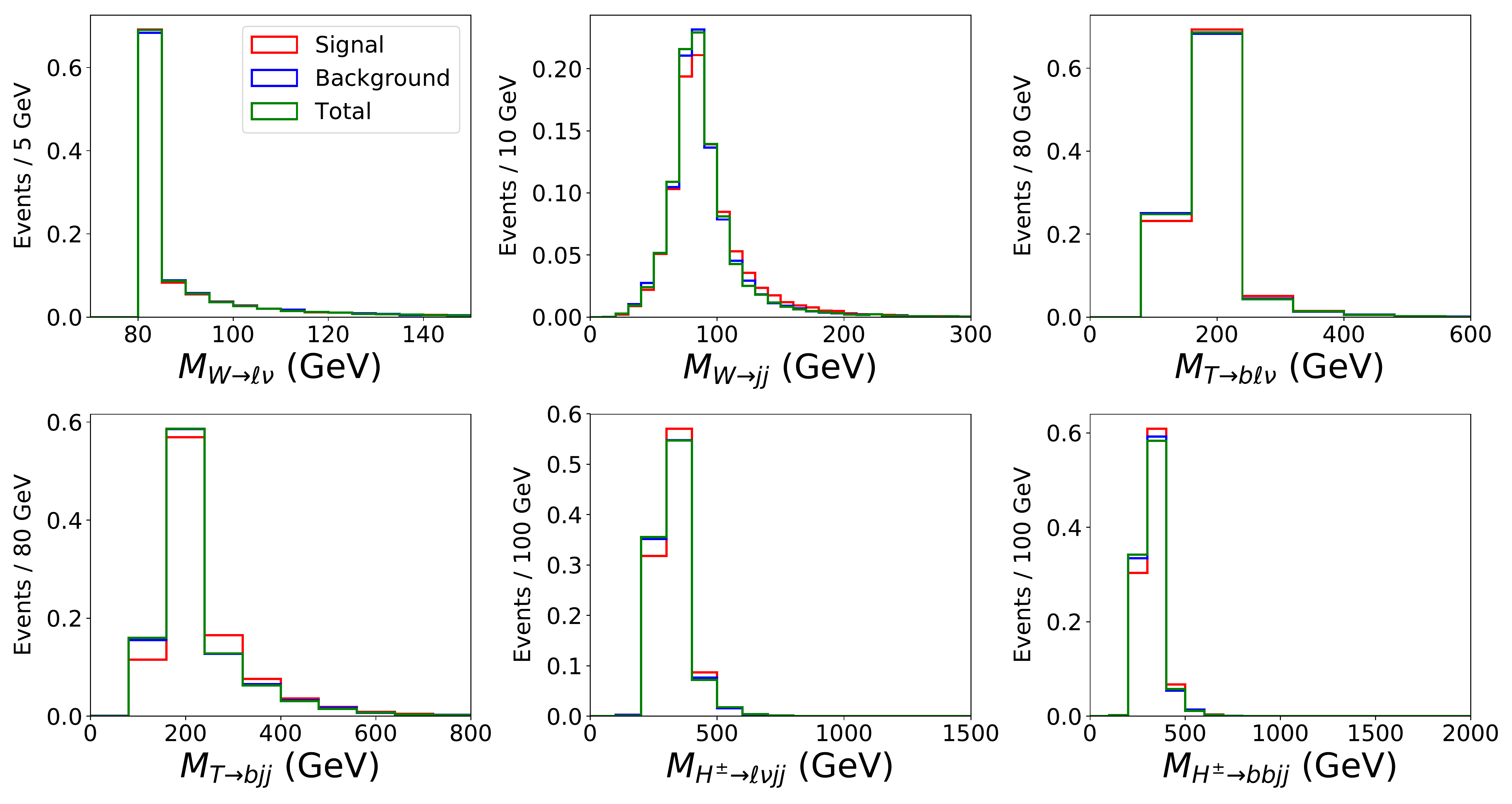}
	\caption{\label{fig:mhmod-recon-variables} Invariant mass distributions for reconstructed particles in the $m_{h}^{\rm mod+}$ benchmark.}
	\end{figure}

The results of this reconstruction can be found in Figs.~\ref{fig:hmssm-recon-variables} and \ref{fig:mhmod-recon-variables}, normalised to unit area.

After this we apply a very simple set of cuts to calculate the behaviour of interference under cutflow. The full set of cuts are exactly one lepton, five or more light jets, two (three) or more b-jets, missing energy greater than $50$ GeV and finally, the transverse mass of missing energy and the lepton must be higher than 60 GeV. Specifically, $m^{W}_{T} = \sqrt{ (\slashed{E}_x + \ell_x)^2 + (\slashed{E}_y + \ell_y)^2} > 60$ GeV. The results of these cutflows can be found in Tabs~\ref{tab:hmssmresults} and \ref{tab:mhmodresults}.

\subsection{The hMSSM analysis}

The parton level interference relative to signal in the hMSSM scenario was $86.4$\%, an alarmingly high level of interference before cuts. The 2 b-tag cut flow served to increase this to $225.6$\% and the 3 b-tag scenario to increase it to $277.8$\%. In all three of these subsets of the phase space the interference appears to be sufficiently large to motivate quantifying it in a full analysis. It should be noted however that the error on the interference is of roughly the same order, and so these results should be considered with this in mind.

\begin{figure}[!ht]
	\centering
	\includegraphics[scale=0.3]{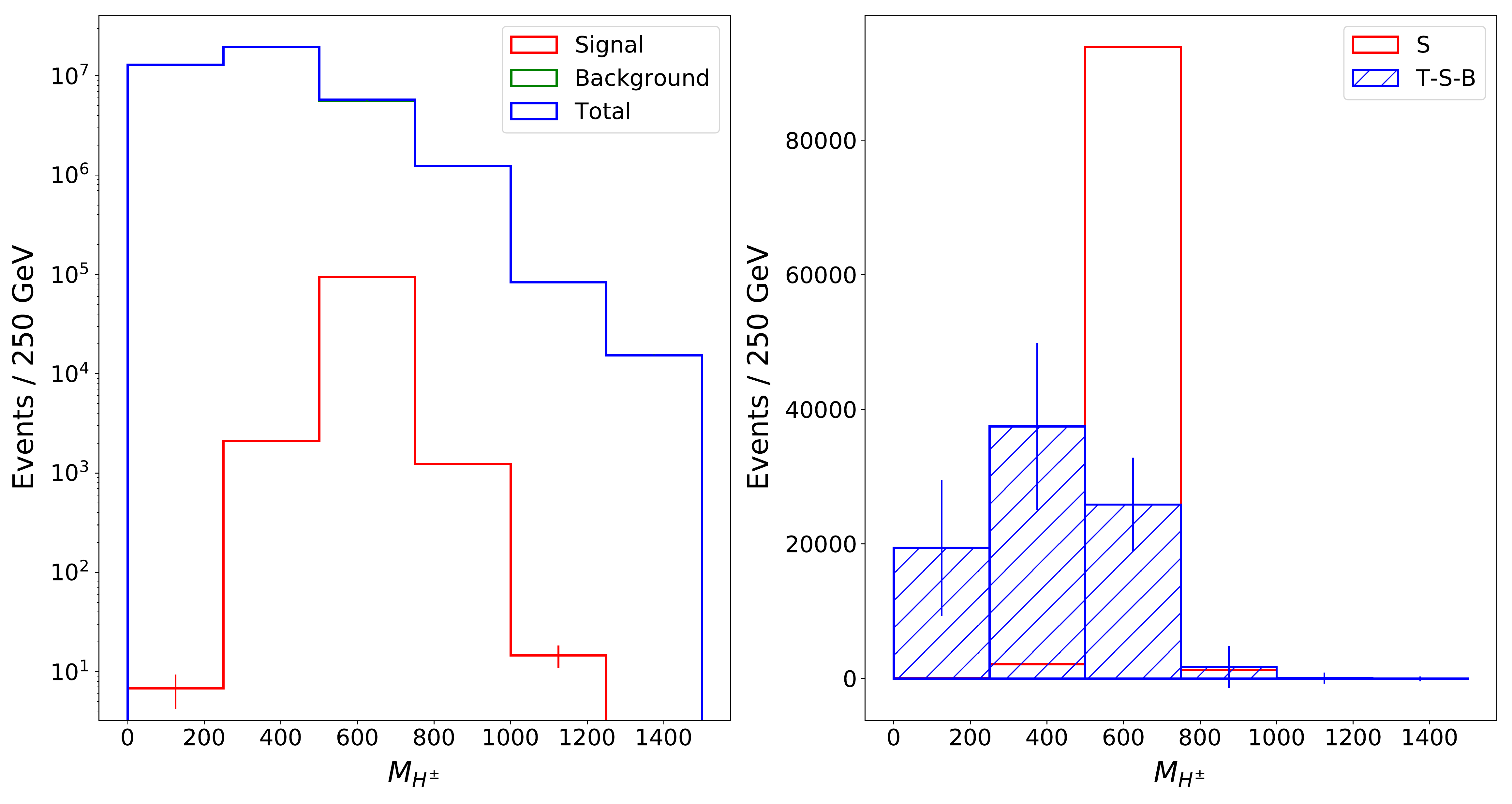}
	\caption{\label{fig:hmssm-interference} The charged Higgs invariant mass distribution  of the signal, background and total samples (left) and interference and signal (right) at parton level {and without cuts} in the $h$MSSM scenario.}
	\end{figure}

Fig~\ref{fig:hmssm-interference} presents an example of the shape of the interference distribution, specifically in the $t\bar{t}b\bar{b}$ reconstructed invariant mass plane. This was undertaken using the large parton level sample so as to ensure the per-bin error was sufficiently low. It can be seen that there is a significant off-peak positive contribution of the interference.

	\begin{table}[!ht]
		\centering
		\begin{tabular}{|lccccc|}
		\hline
	    {Cut} & {S} &  {B} &  {S+B} &  {I} & $\Delta \text{I}$ \\\hline
	    No cuts: &                         97206 &   39235500 &   39417000 &  84294 &  111369 \\\hline
	    $N_{\ell} = 1$:       &            21601 &    9059869 &    9083647 &   2177 &   53488     \\\hline
	    $N_J \geq 5$:       &              19380 &    6256492 &    6296865 &  20991 &   44499     \\\hline
	    $N_{BJ} \geq 2$:       &           15112 &    4058520 &    4091878 &  18246 &   35861      \\\hline
	    $\slashed{E} > 20$ GeV:       &    14356 &    3736396 &    3776148 &  25395 &   34430      \\\hline
	    $\slashed{E} + m_T^W > 60$ GeV: &  14129 &    3639484 &    3685489 &  31874 &   33997     \\\hline\hline
	    {Cut} & {S} &  {B} &  {S+B} &  {I} & $\Delta \text{I}$   \\\hline
	    $N_{BJ} \geq 3$:  &                       8263 &   1715768 & 1733953 &   9921 &   23335\\\hline
	    $\slashed{E} > 20$ GeV:       &           7851 &   1581190 & 1607425 &  18383 &   22435 \\\hline
	    $\slashed{E} + m_T^W > 60$  GeV:       &  7729 &   1540778 & 1569979 &  21471 &   22160\\\hline
		\end{tabular}
		\caption{\label{tab:hmssmresults} Cut flow results presented in expected event yield with 3000 fb$^{-1}$ of luminosity for the $h$MSSM benchmark.}
	\end{table}

\subsection{The $m_{h}^{\rm mod+}$ analysis}

The $m_{h}^{\rm mod+}$ results begin with a smaller parton level interference relative to signal of $15.03$\%, reduce to $3.0$\% in the 2 b-tag scenario and increase to $58.7$\% in the 3 b-tag scenario. However it should be noted that the interference in this case is far smaller than the level of error, and so strong conclusions from this result are not possible.

\begin{figure}[!ht]
	\centering
	\includegraphics[scale=0.3]{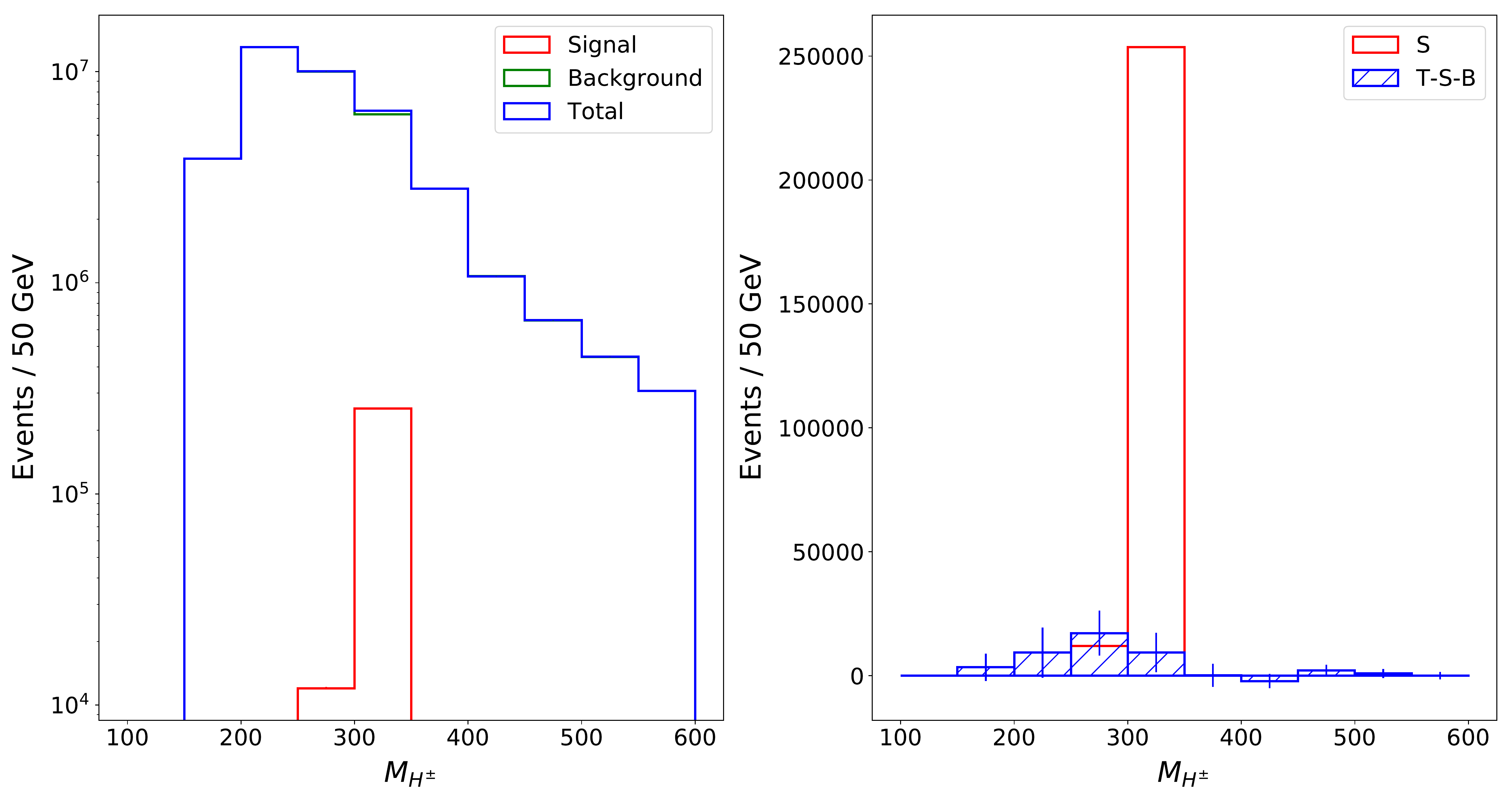}
	\caption{\label{fig:mhmod-interference} The charged Higgs invariant mass distribution  of the signal, background and total samples (left) and interference and signal (right) at parton level {and without cuts} in the $m_{h}^{\rm mod+}$ scenario.}
	\end{figure}

Fig~\ref{fig:mhmod-interference} presents an example of the shape of the interference distribution, again specifically in the $t\bar{t}b\bar{b}$ reconstructed invariant mass plane. This was also undertaken using the large parton level sample and it can be seen that there is a non-negligible positive contribution from interference off-peak.
	
	\begin{table}[!ht]
		\centering
		\begin{tabular}{|lccccc|}
		\hline
	    {Cut} & {S} &  {B} &  {S+B} &  {I} & $\Delta \text{I}$ \\\hline
	    No cuts:  &                        265620 &   39285000 &   39591000 &     40380 &  111923  \\\hline
        $N_{\ell} = 1$:  &                60173 &    9031228 &    9109097 &     17695 &   53673  \\\hline
      	$N_J \geq 5$:   &                 49641 &    6249064 &    6308825 &     10119 &   44671  \\\hline
     	$N_{BJ} \geq 2$:  &                 37040 &    4069533 &    4107962 &      1388 &   36057  \\\hline
    	$\slashed{E} > 20$ GeV:  &          34323 &    3754074 &    3788858 &       460 &   34630  \\\hline
 		$\slashed{E} + m_T^W > 60$ GeV:  &    33422 &    3658612 &    3693048 &      1013 &   34188   \\\hline\hline
        {Cut} & {S} &  {B} &  {S+B} &  {I} & $\Delta \text{I}$ \\\hline
        $N_{BJ} \geq 3$: &                 18946 &   1728147 &   1761007 &    13913 &   23561  \\\hline
        $\slashed{E} > 20$ GeV:  &         17578 &   1594185 &   1626002 &    14238 &   22635  \\\hline
        $\slashed{E} + m_T^W > 60$ GeV:  & 17124 &   1557257 &   1584431 &    10049 &   22357  \\\hline
		\end{tabular}
		\caption{\label{tab:mhmodresults} Cut flow results presented in expected event yield with 3000fb$^{-1}$ of luminosity for the $m_{h}^{\rm mod+}$ benchmark.}
	\end{table}
	
\section{Conclusion}
This talk presented the phenomenological analysis of the size of signal to background interference in the production of $\bar{t}bH^+$ decaying via $H^+ \to t\bar{b}$ and the associated background production of $t\bar{t}b\bar{b}$ at the 13 TeV LHC.

We utilize two benchmarks of the MSSM with heavy charged Higgs, and find that at parton level in both scenarios the interference is non-negligible. Furthermore the shape of the interference distribution and its associated impacts are not necessarily the same as the signal and so a simple rescaling of signal may not be feasible.

Furthermore, while the associated error on the detector level cutflow results is large, the results imply that the interference is sensitive to cuts and can reasonably increase significantly relative to the signal through a simple cutflow.

%
%
%

\end{document}